\documentstyle[12pt]{article}
\newcommand{\eq}{\begin{equation}}
\newcommand{\en}{\end{equation}}
\newcommand{\eqn}{\begin{eqnarray}}
\newcommand{\enn}{\end{eqnarray}}
\newcommand{\nn}{\nonumber \\}
\newcommand{\barr}{\begin{array}}
\newcommand{\earr}{\end{array}}

\newcommand{\NP}{{\it Nucl. Phys.}}
\newcommand{\PL}{{\it Phys.Lett.}}
\newcommand{\CMP}{{\it Comm. Math. Phys.}}
\newcommand{\IJMP}{{\it Int. Journ. Mod. Phys.}}

\newcommand{\A}{\alpha}
\newcommand{\B}{\beta}

\newcommand{\E}{\epsilon}
\newcommand{\DG}{\delta_G}
\newcommand{\DGB}{\bar{\delta}_G}
\newcommand{\AP}{\alpha^{\prime}}
\newcommand{\BP}{\beta^{\prime}}
\newcommand{\APP}{\alpha^{\prime \prime}}
\newcommand{\BPP}{\beta^{\prime \prime}}
\begin{document}
\begin{titlepage}
\begin{flushright}
IASSNS-HEP-92/85 \\
December 1992
\end{flushright}
\vspace{1cm}
\begin{center}
{\LARGE
N=4 Superconformal Algebras and  \\
Gauged WZW Models\\ }

\vspace{1cm}
{\large Murat G\"{u}naydin$^*$} \\
School of Natural Sciences \\
Institute for Advanced Study \\
Olden Lane \\
Princeton, NJ 08540 \\
\vspace{1cm}
{\bf Abstract}
\end{center}
As shown by Witten  the $N=1$ supersymmetric
gauged WZW model based on a group $G$  has an extended $N=2$ supersymmetry if
the gauged subgroup $H$ is so chosen
 that $G/H$ is K\"{a}hler. We extend Witten's
result and
prove  that the $N=1$ supersymmetric gauged WZW models over $G\times U(1)$ are
actually
invariant
under $N=4$ superconformal transformations if the gauged subgroup $H$ is
such  that $G/(H \times SU(2))$ is a quaternionic symmetric space.
 A previous construction of ``maximal''
 $N=4$ superconformal algebras with $SU(2)
\times SU(2) \times U(1)$ symmetry is reformulated and further developed
so as  to relate them to the $N=4$ gauged WZW models. Based on
earlier results we expect the quantization of  $N=4$ gauged WZW models
to yield the unitary realizations of maximal $N=4$ superconformal algebras
provided by this construction.   \\

\vspace{1cm}
\footnoterule
\noindent
{\footnotesize ($^*$) Work supported in part by the
National Science Foundation Grant PHY-9108286. \\
Permanent Address: Physics Department, Penn State University,
University Park, PA 16802 \\
e-mail: Murat@psuphys1.psu.edu or GXT@PSUVM.bitnet}
\end{titlepage}
\setcounter{footnote}{0}
\section{Introduction}
\setcounter{equation}{0}
Wess-Zumino-Witten models \cite{WZ,EW84} have been studied extensively over the
last decade.
They provide  examples of  conformally invariant field theories in two
dimensions and have a
very rich structure \cite{EW84}. Their $N=1$ supersymmetric extensions were
introduced in
\cite{RR} and they exist for any group manifold $G$.
 However, the requirement of
having  more than  one
supersymmetry imposes constraints on the possible WZW models.
This is to be expected from the earlier results on supersymmetric sigma models
and supergravity theories with extended supersymmetry.
For example, in \cite{AF} it was shown
that $N=2$ supersymmetry requires the scalar manifold
of a 2-dimensional sigma model without the Wess-Zumino term  to be Kahler
 and $N=4$ supersymmetry requires it
to be Hyperkahler. The results of Witten \cite{EW84} on
the conformal invariance of WZW models for quantized values of the coefficient
of WZ term provided the motivation for study of supersymmetric sigma models
 with such a term \cite{CZ,GHR,HS}. The supersymmetric sigma models relevant
 for heterotic strings have been studied in \cite{HW,AS}.
 In \cite{SSTP1,SSTP2}
sigma models on group manifolds with extended
rigid supersymmetry were studied
and a classification of all such manifolds with $N=2$ and $N=4$ supersymmetry
was given. More recently  WZW models on the group manifold $SU(2)\times U(1)$
with $N=2$ and $N=4$ supersymmetries were studied using  superspace
techniques \cite{ARSS,IKL}.

In parallel to the work on sigma models a large amount of work has been done
 on the unitary realizations of conformal and superconformal algebras with
a central charge. The Sugawara-Sommerfield contruction \cite{SS}
 was generalized to the coset space construction  of Virasoro algebra
and its $N=1$
supersymmetric extension  in \cite{GKO}.
Kazama and Suzuki showed  that the generalization of the GKO construction to
$N=2$ superconformal algebras requires that the corresponding coset space be
Kahlerian \cite{KS}. The realization of maximal
$N=4$ superconformal algebras with
 $SU(2)\times SU(2)\times U(1)$ symmetry  \cite{STP,KS2} was studied in
\cite{SSTP2,AV}. The required coset spaces for maximal $N=4$ supersymmetry
are of the form $W\times SU(2)\times U(1)$ where $W$ is the quaternionic
symmetric space associated with the group $G$.   In \cite{MG,GH91,GH92}
the unitary realizations of extended superconformal algebras
 ($N=2$ and $N=4$) over triple
systems were given. Only for
 a very special class of triple systems ,called the
Freudenthal triple systems (FTS), the realizations of $N=2$ superconformal
algebras (SCA) admit an extension to $N=4$ SCA's \cite{GH91,GH92}. All the
coset space realizations of $N=4$ SCA's can be thus
obtained from their underlying
FTS's. The unitary representations of the maximal $N=4$ superconformal
algebra were studied in \cite{GPTV} and their characters in \cite{PT}.

Even though the GKO construction and its supersymmetric generalizations
provide us with a large class of unitary representations
 of conformal and superconformal
algebras it is desirable to work with Lagrangian field theories whose
quantization leads to these unitary realizations. The conformally
invariant field theories corresponding to the coset models $G/H$
are the gauged WZW models based on the group $G$ with a gauged
subgroup $H$. This connection was
established  for purely bosonic models in \cite{GWZW,EW92b} and for
 the $N=1$ models in \cite{HS2}. Recently, in his study of matrix models,
 Witten gave a simple formulation
of $N=1$ gauged WZW models in component formalism and extended it to
the $N=2$ supersymmetric theories  corresponding to the Kazama-Suzuki
models \cite{EW92}.
 In this paper we generalize  Witten's results to $N=4$ supersymmetric
gauged WZW models. The first part of the paper is devoted to reformulating
and simplifying the construction of $N=4$ superconformal algebras over
FTS's given in \cite{GH91,GH92} so as to make the connection with the
$N=4$ gauged WZW models more transparent. \footnote{Throughout this paper
the term $N=4$ SCA refers to the maximal $N=4$ superconformal algebra
with $SU(2)\times SU(2) \times U(1)$ symmetry.}

\section{A Construction of $N=2$ Subalgebras of $N=4$ Superconformal
Algebras }
\setcounter{equation}{0}
 As mentioned earlier a general method for the construction of extended
superconformal
algebras over triple systems was developed in \cite{MG,GH91,GH92}.
 For a very special class of triple systems,
namely the Freudenthal triple systems (FTS), the $N=2$ superconformal algebras
thus constructed
can be extended to  $N=4$ superconformal algebras with the symmetry group
$SU(2) \times SU(2)
\times U(1)$ \cite{GH91,GH92}. For FTS's with a non-degenerate symplectic form
these
realizations of $N=2$ superconformal algebras are equivalent to their
realization over
certain Kahlerian coset spaces of Lie groups a la Kazama and Suzuki \cite{KS}.
For a simple
group $G$ the associated coset space is $G/H\times U(1)$ where $H$ is such that
$G/H\times SU(2)$ is a quaternionic symmetric space as expected from the
results of
\cite{SSTP2,GPTV}. We shall now review and reformulate  this construction in a
way that does not require
familiarity with the underlying FTS's.

Let $g,h$ be the Lie algebras of $G$ and $H$, respectively.
 The Lie algebra $g$ can be given a 5-graded
structure
\eq
g = g^{-2} \oplus g^{-1} \oplus g^{0} \oplus g^{+1} \oplus g^{+2}
\en
such that $g^{0}=h \oplus K_{3}$ where $K_{3}$ is the generator of the $U(1)$
factor and the
grade $\pm 2$ subspaces are one dimensional. Let us denote them as
\eqn
K_{+}&=&K_{1}+iK_{2} \in g^{+2} \nn
&&   \nn
K^{+}&=&K_{-}= K_{1}-iK_{2} \in g^{-2}
\enn
The elements of the grade $+1$ and $-1$ subspaces will be denoted as $U_{a}$
and
$U^{a}$, respectively, where $a,b,..=1,2,..,D$ with $D$ being the dimension of
the underlying
FTS. There is a universal relation between $D$ and the dual Coxeter number
$\check{g}$ of $G$:
\eq
D=2(\check{g}-2)
\en
 Under hermitian conjugation we have
\eqn
K_{-}& =& K_{+}^{\dagger}\nn
&&  \nn
U^{a}& =& U_{a}^{\dagger}
\enn
The generators $K_{-}, K_{+}$ and $K_{3}$ form an $SU(2)$ subalgebra of $g$.
\eqn
{[K_{+},K_{-}]}& =&2K_{3} \nn
&&  \nn
{[K_{3}, K_{\pm}]}& =& \pm K_{\pm}
\enn
The commutation relations of the U's are
\eqn
{[U_{a},U_{b}]}& =& \Omega_{ab} K_{+} \nn
&&  \nn
{[U^{a}, U^{b}]}& =& \Omega^{ab} K_{-} \nn
&&  \nn
{[U_{a}, U^{b}]}& =& S_{a}^{b}
\enn
where $\Omega_{ab}$ is a symplectic invariant tensor of $H$ and $S_{a}^{b}$ are
the
generators of the subgroup $H\times U(1)$.  The tensor $\Omega^{ab}$ is the
inverse of
$\Omega_{ab}$ and satisfies :
\eqn
\Omega_{ab} \Omega^{bc}& =& \delta_{a}^{c} \nn
&&  \nn
\Omega_{ab}^{\dagger}& =& \Omega^{ba} =-\Omega^{ab}
\enn
The trace component of $S_{a}^{b}$ gives the $U(1)$ generator
\eq
K_{3} =\frac{1}{2(\check{g}-2)} S_{a}^{a}
\en
Therefore we have the decomposition
\eq
S_{a}^{b} = H_{a}^{b} + \delta_{a}^{b} K_{3}
\en
where $H_{a}^{b} = S_{a}^{b} - \frac{1}{D} \delta_{a}^{b} S_{c}^{c}$ are the
generators of the subgroup $H$. Note that $H_{a}^{b}$ commutes with $K_{3},
K_{+} $ and $K_{-}$.
 The other non-vanishing commutators of $g$ are
\eqn
{[K_{+}, U^{a}]}& =& \Omega^{ab} U_{b} \nn
&&  \nn
{[K_{-}, U_{a}]}& =& \Omega_{ab} U^{b} \nn
&& \nn
{[K_{3}, U^{a}]}& =& - \frac{1}{2} U^{a} \nn
&&  \nn
{[K_{3}, U_{a}]}& =& \frac{1}{2} U_{a} \nn
&& \nn
{[S_{a}^{b}, U_{c}]}& =& \Sigma_{ac}^{bd} U_{d} \nn
&& \nn
{[S_{a}^{b}, U^{c}]}& =& -\Sigma_{ad}^{bc} U^{d} \nn
&&  \nn
{[S_{a}^{b}, S_{c}^{d}]}& =& \Sigma_{ac}^{be} S_{e}^{d} -\Sigma_{ae}^{bd}
S_{c}^{e}
\enn
where $\Sigma_{ab}^{cd}$ are the structure constants of the corresponding FTS
which are normalized such that
\eqn
\Sigma_{ab}^{ac}& =& (\check{g} -2) \delta_{b}^{c} \nn
&& \nn
\Sigma_{ab}^{bc}& =& (\check{g} -1) \delta_{a}^{c}  \nn
&& \nn
\Sigma_{ab}^{cd} - \Sigma_{ab}^{dc}& =& \Omega_{ab} \Omega^{cd}
\enn
Consider now the affine Lie algebra $\hat{g}$ defined by $g$. It can similarly
be given a 5-graded structure with
the central charge belonging to the grade zero subspace. The
commutation relations of
$\hat{g}$ can be written in the form of operator products as follows
\cite{GH91,GH92}:
\begin{equation}
\begin{array}{lll}

U_{a}(z) U^{b}(w) & = &
 \frac{k \delta_{a}^{b}}{(z - w)^{2}} + \frac
{S_{a}^{b}(w)}{(z - w)} + \cdots \\
&  &   \\
U_{a}(z) U_{b}(w) & = &
 \frac{\Omega_{ab} K_{+}(w)}{(z - w)} + \cdots \\
&  &   \\

S_{a}^{b}(z) U_{c}(w) & = &
 \frac{\Sigma_{ac}^{bd} U_{d}(w)}{(z-w)} + \cdots \\
&  &   \\

S_{a}^{b}(z) S_{c}^{d}(w) & = &
 \frac{k \Sigma_{ac}^{bd}}{(z-w)^{2}}
 + \frac{1}{(z-w)} (\Sigma_{ac}^{be}S_{e}^{d} - \Sigma_{ae}^{bd} S_{c}^{e})(w)
+ \cdots \\
&  &   \\
K_{3}(z) K_{\pm}(w) & =  &
  \frac{\pm K_{\pm}(w)}{(z-w)} + \cdots \\
&  &  \\
K_{+}(z) K_{-}(w) &= & \frac{k}{(z-w)^{2}} +\frac{2J_{3}(w)}{(z-w)}+ \cdots \\
\ & \ &  \\
K_{+}(z) U^{a}(w) & = & \frac{\Omega^{ab}U_{b}(w)}{(z-w)}
 + \cdots  \\
&  &   \\
K_{-}(z) U_{a}(w) & = & \frac{\Omega_{ab} U^{b}(w)}{(z-w)}
 + \cdots

\end{array}
\end{equation}

To construct the supersymmetry generators we also need to introduce fermi
fields corresponding to the coset space $G/H\times U(1)$ \cite{GH91,GH92}:

\begin{equation}
\begin{array}{l}

\psi_{a}(z) \psi^{b}(w) = \frac{\delta_{a}^{b}}{(z - w)} + \cdots  \\
\ \\
\psi_{+}(z) \psi^{+}(w) = \frac{1}{(z - w)} + \cdots  \\
\ \\
\psi_{a}(z) \psi_{b}(w) = \cdots  \\
\ \\
\psi_{+}(z) \psi_{+}(w) = \cdots  \\
\ \\
\psi_{a}(z) \psi^{+}(w) = \cdots

\end{array}
\end{equation}

Then the supersymmetry generators are given by the following expressions
\footnote{Composite operators with a single argument are assumed to be
normal ordered.}
\begin{equation}
\begin{array}{lll}

G(z) & = & \sqrt{\frac{2}{k+ \check{g}}}
 \{ U_{a} \psi^{a} + K_{+} \psi^{+} -
\frac{1}{2} \Omega_{ab} \psi^{a} \psi^{b} \psi_{+} \}(z) \\
\end{array}
\end{equation}
\begin{equation}
\begin{array}{lll}
\bar{G}(z) & = & \sqrt{\frac{2}{k+ \check{g}}}
 \{ U^{a} \psi_{a} +
 K^{+} \psi_{+} -
 \frac{1}{2} \Omega^{ab} \psi_{a} \psi_{b} \psi^{+} \}(z)  \\
\end{array}
\end{equation}
They satisfy

\begin{equation}
\begin{array}{l}

G(z)\bar{G}(w) = \frac{2c/3}{(z-w)^{3}} + \frac{2J(w)}{(z-w)^{2}}
+ \frac{2T(w) + \partial J(w)}{(z-w)} + \cdots \\
\ \\
T(z)T(w) = \frac{c/2}{(z-w)^{4}} + \frac{2T(w)}{(z-w)^{2}}
+ \frac{\partial T(w)}{(z-w)} + \cdots \\
\ \\
J(z)G(w) = \frac{G(w)}{(z-w)} + \cdots \\
\ \\
J(z)\bar{G}(w) = - \frac{\bar{G}(w)}{(z-w)} + \cdots \\
\ \\
T(z)J(w) = \frac{J(w)}{(z-w)^{2}} + \frac{\partial J(w)}{(z-w)}+\cdots \\
\ \\
T(z)G(w) = \frac{\frac{3}{2}G(w)}{(z-w)^{2}} +
 \frac{\partial G(w)}{(z-w)}+ \cdots \\
\ \\
T(z)\bar{G}(w) = \frac{\frac{3}{2} \bar{G}(w)}{(z-w)^{2}} +
 \frac{\partial \bar{G}(w)}{(z-w)} + \cdots
\end{array}
\end{equation}
where

\begin{equation}
\begin{array}{lll}
T(z) & = & \frac{1}{k+ \check{g}}
\{ \frac{1}{2}(U_{a}U^{a} + U^{a}U_{a})
+ \frac{1}{2}(K_{+}K^{+} + K^{+}K_{+})   \\
     &  &  \\
     &   & - \frac{k+1}{2}(\psi_{a} \partial \psi^{a} +
\psi^{a} \partial \psi_{a})
- \frac{1}{2}(k+ \check{g}-2)(\psi_{+} \partial \psi^{+} +
\psi^{+} \partial \psi_{+})   \\
     &  &  \\
      &   & + H_{a}^{b} \psi^{a} \psi_{b} +
K_{3}(\psi^{a} \psi_{a} +2 \psi^{+} \psi_{+}) +
       \psi_{+} \psi^{+} \psi^{a} \psi_{a} +
\frac{1}{4}\Omega_{ab} \psi^{a} \psi^{b} \Omega^{cd}
 \psi_{c} \psi_{d} \}(z)
\end{array}
\end{equation}
\vspace{1.0cm}
\begin{equation}
J(z) = \frac{1}{k+ \check{g}} \{ 2(\check{g} - 1) K_{3}
 + (k+1)\psi^{a}
 \psi_{a} + (k- \check{g}+2) \psi^{+} \psi_{+} \}(z)
\end{equation}

The above realization of $N=2$ SCA's corresponds to the coset $G/H\times U(1)$
where
the $U(1)$ generator $K_{3}$  determines the 5-graded structure of $g$.
Formally, one can write
\eq
{[2K_{3}, g^{m}]} = m g^{m}
\en
where $g^{m}$ denotes the subspace of grade $m $ with
$m=0,\pm1,\pm2 $.

The Lie algebra $g$  can be given a 5-graded structure with respect to $K_{1}$
as well as $K_{2}$.
Therefore, one can realize the $N=2$ SCA equivalently over the coset $G/H\times
U(1)^{\prime} $
or the coset $G/H\times U(1) ^{\prime\prime}$, where the generators of
$U(1)^{\prime}$ and
$U(1)^{\prime\prime}$ are $K_{1}$ and $K_{2}$, respectively. The grade $\pm1$
and $\pm2$ subspaces
with respect to $K_{1}$ are
\eqn
U_{a}^{\prime}& =& \frac{1}{\sqrt{2}} ( U_{a} + \Omega_{ab} U^{b})\nn
&& \nn
U^{a^{\prime}}& =& \frac{1}{\sqrt{2}} ( U^{a} - \Omega^{ab} U_{b} ) \nn
&&  \nn
K_{+}^{\prime}& =& i(K_{1}+iK_{2}) \nn
&&  \nn
K_{-}^{\prime}& =& -i (K_{1}-iK_{2})
\enn
They satisfy
\eqn
{[U_{a}^{\prime},U_{b}^{\prime} ]}& =& \Omega_{ab} K_{+}^{\prime} \nn
&& \nn
{[U^{a^{\prime}},U^{b^{\prime}}]}&=& \Omega^{ab} K_{-}^{\prime} \nn
&&  \nn
{[K_{+}^{\prime}, U^{a^{\prime}} ]}& =& \Omega^{ab} U_{b}^{\prime} \nn
&&  \nn
{[K_{-}^{\prime}, U_{a}^{\prime}]}& = & \Omega_{ab} U^{b^{\prime}}
       \label{eq:UP}
\enn
The grade $\pm1$ and $\pm2$ subspaces with respect to $K_{2}$ are
\eqn
U_{a}^{\prime\prime}& =& \frac{1}{\sqrt{2}} (U_{a} +i \Omega_{ab} U^{b})\nn
&&  \nn
U^{a^{\prime\prime}}& =& \frac{1}{\sqrt{2}} (U^{a} +i\Omega^{ab} U_{b} ) \nn
&& \nn
K_{+}^{\prime\prime}& =& -i(K_{3} +iK_{1}) \nn
&&  \nn
K_{-}^{\prime\prime}& =& i(K_{3} -iK_{1})
 \enn
with the commutation relations similar to those of equations (\ref{eq:UP}).

For every simple Lie group $G$ (except for $SU(2)$) there exists a subgroup
$H \times U(1)$
, unique up to automorphisms, such that $g$ has a 5-graded structure with
respect
to the subalgebra $h \oplus K$, where $K$ is the generator of the $U(1)$
subgroup.
Below we list the simple groups $G$ together with their subgroups $H$:
\begin{center}
\begin{tabular}{|c|c|}  \hline
$G$ & $H$ \\
\hline
$SU(n)$ & $U(n-2)$ \\
$SO(n)$ & $SO(n-4) \times SU(2)$ \\
$Sp(2n)$ & $Sp(2n-2)$  \\
$G_{2}$ & $SU(2)$ \\
$F_{4}$ & $Sp(6)$ \\
$E_{6}$ & $SU(6)$ \\
$E_{7}$ & $SO(12)$ \\
$E_{8}$ & $E_{7}$ \\
\hline
\end{tabular}
\end{center}

\section{The Construction of $N=4$ Superconformal Algebras}
\setcounter{equation}{0}
The commutation relations of
the $N=4$ SCA, denoted as  $\cal{A_{\gamma}}$,
  with the gauge group $SU(2)\times SU(2) \times U(1)$ and four dimension $1/2$
generators  can be written as the following operator products
relations \cite{STP}\footnote{Note that our conventions and normalizations
differ
from those of references \cite{SSTP2,STP,GPTV,AV}. The generators
$(G_a,T,V^{\pm i}, Z ,\xi^a)$ appearing above are all
hermitian operators.}
\begin{eqnarray}
G_{a}(z)G_{b}(w)&=&\frac{2c}{3}\delta_{ab}(z-w)^{-3}+(z-w)^{-2}2M_{ab}(w)
\nonumber\\
&&+(z-w)^{-1}[2T(w)\delta_{ab}+\partial M_{ab}(w)]+\cdots \nonumber\\
M_{ab}&=&\frac{4}{(k^{+}+k^{-})} [k^{-}\alpha^{+i}_{ab}V^{+}_{i} + k^{+}
\alpha^{-i}_{ab}V^{-}_{i}]\nonumber\\
V^{\pm i}(z) G_{a}(w)&=&\alpha^{\pm i\ b}_{\ \ a} [G_{b}(w)\ (z-w)^{-1}
\mp \frac{2}{k^{+}+k^{-}}k^{\pm}\,\xi_{b}(w)\ (z-w)^{-2}]+\cdots\nonumber\\
V^{\pm i}(z)\,V^{\pm j}(w)&=& i\varepsilon^{ijk}V^{\pm_k}(w)\ (z-w)^{-1} -
\frac{k^\pm}{2}\delta^{ij}(z-w)^{-2}+\cdots\nonumber\\
\xi_{a}(z)\,G_{b}(w)&=&\left[
\frac{-2i}{\sqrt{k^{+}+k^{-}}}(\alpha^{+i}_{ab}V^{+}_{i}(w)\ - \alpha^{-i}_{ab}
V^{-}_{i}(w)) - \frac{\delta_{ab}}{\sqrt{2}}Z(w)\right]
(z-w)^{-1}+\cdots\nonumber\\
V^{\pm i}(z)\,\xi_{a}(w)&=&\alpha^{\pm i\ b}_{\ \ a}\xi_{b}(w)\
(z-w)^{-1}\nonumber\\
Z(z)\,G_{a}(w)&=& -\sqrt{2}\xi_{a}(z-w)^{-2}+\cdots \nn
\xi_{a}(z)\,\xi_{b}(w)&=& \frac{1}{2}\delta_{ab}(z-w)^{-1}+\cdots\nonumber\\
Z(z)\,Z(w)&=& (z-w)^{-2}+\cdots  \nn
 a,b,.. &=&1,2,3,4  \nn
 i,j,..& =&1,2,3
\end{eqnarray}
plus the usual operator products of the Virasoro generator $T(z)$ with itself
and the other
generators.
The $\alpha^{\pm i} $ are $4\times 4$ matrices satisfying
\eqn
{[}\alpha^{\pm i}, \alpha^{\pm j}{]}& =& i \epsilon^{ijk} \alpha^{\pm}_k \nn
{[}\alpha^{+i}, \alpha^{-j}{]}&=&0  \nn
\{\alpha^{\pm i}, \alpha^{\pm j} \}& =& \frac{\delta^{ij}}{2}
\enn

The $V^{\pm i}(z)$ are the currents of the $SU(2)^{+}\times SU(2)^{-}$ symmetry
and
$Z(z)$ is the $U(1)$ current. The $\xi_{a}$ are the four dimension $1/2$
generators of the
algebra. The central charge of the $N=4$ SCA is simply
\eq
c= \frac{6k^{+}k^{-}}{k^{+}+k^{-}}
\en
where $k^{\pm}$ are the levels of the two $SU(2)$ currents \cite{SSTP2}.

There are many different ways of truncating the $N=4$ SCA to an $N=2$ SCA. For
example,
any pair of operators $ (G_{a}+iG_{b})$ and $(G_{a}-iG_{b}) $ generate an $N=2$
SCA.
 A "maximal"
$N=2$ supersymmetric truncation leads to a $N=2$ SCA in semidirect sum with an
$N=2$ " matter
multiplet "\cite{GPTV}. The matter multiplet consists of a complex  current
$A(z)$ and a complex
fermion $Q$ of dimensions 1 and 1/2, respectively.
 The generators of the $N=2$ SCA appearing in such a truncation
can be decomposed as a direct sum \cite{GPTV}
\eqn
T(z)&=& \hat{T}(z) + T_{Q}(z) \nn
&& \nn
G(z)&=& \hat{G}(z) + G_{Q}(z) \nn
&& \nn
\bar{G}(z)& =& \hat{\bar{G}}(z) + \bar{G}_{Q}(z) \nn
&& \nn
J(z)&=& \hat{J}(z) + J_{Q}(z)
\enn
where $T_Q, G_Q, \bar{G}_Q $ and $J_Q$ are bilinears of the
matter multiplet
\eqn
T_{Q}& =& \frac{1}{2} ( A  A^{*} + \partial Q  Q^{*} + \partial Q^{*} Q )\nn
&&  \nn
G_{Q}& =& A^{*} Q \nn
&& \nn
\bar{G}_{Q}& =& A Q^{*} \nn
\nn
J_{Q}& =& Q Q^{*}
 \enn
The operator product of $\hat{T}, \hat{G}, \hat{J}$ with $T_{Q},G_{Q},J_{Q}$
are regular.
The "irreducible" realizations of the $N=2$ SCA generated by $\hat{T}, \hat{G},
\hat{\bar{G}}$ and
$\hat{J}$ over the coset spaces of simple Lie groups can all be obtained by the
construction
outlined in the previous section.
To extend the $N=2$ SCA's of the previous section to $N=4$ SCA's one needs to
introduce a matter
multiplet  and define two additional supersymmetry generators  as well as
adding the matter contribution to the first two
 supersymmetry generators \cite{GH91,GH92}.
The  required currents of the matter multiplet turn out to be the $U(1)$
current generated by $K_{3}$ that gives the
5-graded structure of the Lie algebra $g$, and an additional $U(1)$ current
whose
generator $K_{0}$ commutes with $g$ together with the associated fermions
which we denote
as a complex fermion $\chi_{+}$ and its conjugate $\chi^{+}$. Then
the four supersymmetry  generators of the $N=4$ SCA can be written as

\begin{equation}
\begin{array}{lll}

\frac{1}{\sqrt{2}} (G_{1}+ i G_{2}) & = & \sqrt{\frac{2}{k+\check{g}}}
\{ U_{a} \psi^{a} + K_{+} \psi^{+} +
 K_{3} \chi_{+}  \\
  &  &  \\
&   & - \frac{1}{2} \Omega_{ab} \psi^{a} \psi^{b} \psi_{+} -
\frac{1}{2} \psi^{a} \psi_{a} \chi_{+} - \psi^{+} \psi_{+} \chi_{+} \}
+ iZ \chi_{+}  \\
  &  &  \\
\frac{1}{\sqrt{2}} (G_{1}- i G_{2}) & = & \sqrt{\frac{2}{k+\check{g}}}
\{ U^{a} \psi_{a} + K^{+} \psi_{+}
+ K_{3} \chi_{+} \\
 &  &  \\
&  & - \frac{1}{2} \Omega^{ab} \psi_{a} \psi_{b} \psi^{+} -
\frac{1}{2}\psi^{a} \psi_{a} \chi^{+} - \psi^{+} \psi_{+} \chi^{+} \}
-i Z \chi^{+}  \\
 &  &  \\
\frac{1}{\sqrt{2}}(G_{3}+ i G_{4}) & = & \sqrt{\frac{2}{k+\check{g}}}
\{ \Omega^{ab} U_{a} \psi_{b} + K_{+} \chi^{+} +
K_{3} \psi_{+}  \\
&  &  \\
&   & + \frac{1}{2} \Omega^{ab} \psi_{a} \psi_{b} \chi_{+} +
\frac{1}{2} \psi^{a} \psi_{a} \psi_{+} - \chi^{+} \chi_{+} \psi_{+} \}
+ iZ \psi_{+}  \\
&  &  \\
\frac{1}{\sqrt{2}}(G_{3}- i G_{4}) & = & \sqrt{\frac{2}{k+\check{g}}}
\{ \Omega_{ab} U^{b} \psi^{a} + K^{+} \chi_{+} +
K_{3} \psi^{+}  \\
&  &  \\
&   & + \frac{1}{2} \Omega_{ab} \psi^{a} \psi^{b} \chi^{+} +
\frac{1}{2} \psi^{a} \psi_{a} \psi^{+} - \chi^{+} \chi_{+} \psi^{+} \}
-i Z \psi^{+}  \\

\end{array}
\end{equation}

The Virasoro generator of ${\cal A}_{\gamma}$  is given by
\begin{equation}
\begin{array}{lll}
T(z) & = & \frac{1}{2} \left[ Z^{2} -
 (\chi_{+} \partial \chi^{+} +
\chi^{+} \partial \chi_{+}) - (\psi_{+} \partial \psi^{+} +
\psi^{+} \partial \psi_{+}) \right](z)   \\
& &  \\
     &   & + \frac{1}{k+\check{g}} \{\frac{1}{2}(U_{a}U^{a} +
U^{a}U_{a})
+ \frac{1}{2}(K_{+}K^{+} + K^{+}K_{+}) + K_{3}^{2} \\
&  &  \\
     &   & - \frac{k+1}{2}(\psi_{a} \partial \psi^{a} +
\psi^{a} \partial \psi_{a}) +H_{a}^{b} \psi^{a} \psi_{b} +
\frac{1}{4}\Omega_{ab}
 \psi^{a} \psi^{b} \Omega^{cd} \psi_{c} \psi_{d} \}(z)
\end{array}
\end{equation}

The generators of the two $SU(2)$  currents take the form
\begin{equation}
\begin{array}{lll}

V_{3}^{+}(z) & = & K_3(z) +
\frac{1}{2}( \psi_{+} \psi^{+} + \chi_{+} \chi^{+})(z) \\
&  &  \\
V_{+}^{+}(z) & = & (V_{1}^{+}+ i V_{2}^{+})(z) =
(K_{+} - \psi_{+} \chi_{+})(z) \\
& &  \\
V_{-}^{+}(z) & = &(V_{1}^{+}- i V_{2}^{+})(z) =
(K^{+} + \psi^{+} \chi^{+})(z) \\
&  &  \\
V_{3}^{-}(z) & = & \frac{1}{2}(\psi^{a}\psi_{a} +
\psi^{+}\psi_{+} + \chi_{+}\chi^{+} )(z) \\
& & \\
V_{+}^{-}(z) & = & (V_{1}^{-}+ i V_{2}^{-})(z) =
(\psi^{+} \chi_{+} -
 \frac{1}{2}\Omega_{ab} \psi^{a} \psi^{b})(z)   \\
&  &  \\
V_{-}^{-}(z) & = & (V_{1}^{-}- i V_{2}^{-})(z) =
(\chi^{+} \psi_{+} -
 \frac{1}{2}\Omega^{ab} \psi_{a} \psi_{b})(z)

\end{array}
\end{equation}
The  U(1) current of the $N=4$ SCA is  $Z(z)$ and the  four
dimension $\frac{1}{2}$ generators  are simply the
fermion fields $\psi_{+}(z), \psi^{+}(z), \chi_{+}(z)$ and $\chi^{+}(z)$.
One finds that the levels of the two $SU(2)$ currents are \cite{GH91,GH92}
\begin{equation}
\begin{array}{l}
k^{+}= k+1 \\
\  \\
k^{-}= \check{g} - 1
\end{array}
\end{equation}
where $k$ is the level of $\hat{g}$. The above realization of the $N=4$ SCA
corresponds to the coset
space $G \times U(1)/H$.

Interestingly. one can decouple the four dimension $\frac{1}{2}$  operators and
the $U(1)$ current
$Z(z)$ so as to obtain a non-linear $N=4$ SCA \cite{GPTV,GS} a la Bershadsky
and Knizhnik \cite{MB,VK}.
The generators of this non-linear algebra take the form
\eqn
 \tilde{T} &=& T -(\frac{1}{2} ZZ + \partial \xi^{a} \xi_{a})\nn
\tilde G_{a} &=& G_{a} + \sqrt{2} Z\xi_{a} -\frac{2i}{3\sqrt{(k^{+}+k^{-})}}
\epsilon_{abcd} \xi^{b} \xi^{c} \xi^{d}
+\frac{4i}{\sqrt{k^{+}+k^{-}}}\xi^{b}(
\alpha^{+i}_{ba}\tilde V^{+}_{i}
-\alpha^{-i}_{ba}\tilde V^{-}_{i}) \nn
\tilde V^{\pm i}&=& V^{\pm i}+\alpha^{\pm i}_{ab}\xi^{a}\xi^{b}
\enn
where the conventions for the epsilon symbol are determined by the relation
\eq
\alpha^{\pm iab} \alpha^{\pm \ cd}_{i} = \frac{1}{4} ( \delta^{ac}\delta^{bd} -
\delta^{ad} \delta^{bc}
\pm \epsilon^{abcd} )
\en
They satisfy the operator product relations
\eqn
\tilde{V}^{\pm i}(z) \tilde{G}_{a}(w)&=&
\alpha^{\pm i}_{ab}\,\tilde{G}^{b}(w) (z-w)^{-1}+\ldots\nn
\tilde{G}_{a}(z)\tilde{G}_{b}(w)&=&\frac{4\tilde{k}^+\tilde
k^-}{k^{+}+k^{-}}\delta_{ab}
(z-w)^{-3}+2\tilde T\ \delta_{ab}(z-w)^{-1}\nn
&&+\frac{8}{k}\left(\tilde k^{-}\alpha^{+i}_{ab}\tilde V^{+}_{i}
+\tilde k^{+}\alpha^{-i}_{ab}\tilde V^{-}_{i}\right)(z-w)^{-2}\nn
&&+\frac{4}{k}\partial\left(\tilde k^-\alpha^{+i}_{ab}\tilde V^+_i
+\tilde k^+\alpha^{-i}_{ab}\tilde V^-_i\right)(z-w)^{-1}\\
&&-\frac{8}{k}
(\alpha^{+i}\tilde V^+_i-\alpha^{-i}\tilde V^-_i)_{c(a}
(\alpha^{+j}\tilde V^+_j-\alpha^{-j}\tilde V^-_j)_{b)}^{\ \ c}
(z-w)^{-1}.\nonumber
\enn
where $(a \cdots b)$ in the last expression means symmetrization with respect
to $a$ and $b$.

Note that the supersymmetry generators close into bilinears of currents and the
levels of the
two  $SU(2)$ currents $\tilde{V}^{\pm}$ are $\tilde{k}^{\pm} = k^{\pm}-1 $. The
central charge
is
\eq
\tilde{c} = c-3
\en
The realization given above for the $N=4$ SCA leads to very simple expressions
for the
supersymmetry generators  of the non-linear algebra
\eqn
\frac{1}{\sqrt{2}} (\tilde{G}_{1}+i\tilde{G}_{2})& =&
\sqrt{\frac{2}{k+\check{g}}} U_{a} \psi^{a} \nn
\frac{1}{\sqrt{2}} (\tilde{G}_{1}-i\tilde{G}_{2})& =&
\sqrt{\frac{2}{k+\check{g}}} U^{a} \psi_{a} \nn
\frac{1}{\sqrt{2}} (\tilde{G}_{3}+i\tilde{G}_{4})&=&
\sqrt{\frac{2}{k+\check{g}}} \Omega^{ab} U_{a} \psi_{b}   \nn
\frac{1}{\sqrt{2}} (\tilde{G}_{3}-i\tilde{G}_{4})& =&
\sqrt{\frac{2}{k+\check{g}}} \Omega_{ba} U^{a} \psi^{b}
\enn
The generators of the two $SU(2)$ currents also simplify
\eqn
\tilde{V}^+_+& =& K_+ \nn
\tilde{V}^+_-& =& K_- \nn
\tilde{V}_3& =& K_3  \nn
\tilde{V}^-_+& =& -\frac{1}{2} \Omega_{ab} \psi^a \psi^b \nn
\tilde{V}^-_-& =& - \frac{1}{2} \Omega^{ab} \psi_a \psi_b  \nn
\tilde{V}^-_3& =& \frac{1}{2} \psi^a \psi_a
\enn
The Virasoro generator of the nonlinear algebra is then
\begin{equation}
\begin{array}{lll}
T(z) & = &\frac{1}{k+\check{g}} \{ \frac{1}{2}(U_{a}U^{a} +
U^{a}U_{a})
+ \frac{1}{2}(K_{+}K^{+} + K^{+}K_{+}) + K_{3}^{2} \\
&  &  \\
     &   & - \frac{k+1}{2}(\psi_{a} \partial \psi^{a} +
\psi^{a} \partial \psi_{a}) +H_{a}^{b} \psi^{a} \psi_{b} +
\frac{1}{4}\Omega_{ab}
 \psi^{a} \psi^{b} \Omega^{cd} \psi_{c} \psi_{d} \}(z)
\end{array}
\end{equation}
It is clear from the above expressions for the generators that the non-linear
$N=4$  SCA is
realized over the symmetric space
\eq
G/H\times SU(2)
\en
which is the unique quaternionic symmetric space associated with $G$ .

\section{Witten's Formulation of $N=1$ and $N=2$ Supersymmetric Gauged WZW
Models}
\setcounter{equation}{0}
The $N=1$ supersymmetric WZW models were studied in \cite{RR,SSTP1} and  their
gauged
versions in \cite{HS2}. More recently, Witten gave a simple formulation of the
$N=1$ supersymmetric WZW
and gauged WZW models and generalized it  to models with  $N=2$
supersymmetry \cite{EW92}. In this section we
  review Witten`s construction which we shall generalize to $N=4$ gauged
WZW models in the next section.

The WZW action at level $k$ is given by $kI(g)$ where
\eq
I(g) = -\frac{1}{8\pi} \int_{\Sigma} d^2 \sigma \sqrt{h} h^{ij}
Tr(g^{-1}\partial_ig \cdot g^{-1}
\partial_j g) -i \Gamma
\en
with the WZ functional \cite{WZ} given by \cite{EW84}
\eq
\Gamma = \frac{1}{12\pi} \int_M d^3 \sigma \epsilon^{ijk} Tr ( g^{-1}
\partial_ig \cdot
g^{-1} \partial_jg \cdot g^{-1} \partial_k g)
\en
 $M$ is any three manifold whose boundary is the Riemann surface $\Sigma$
with metric $h$. $g$  is a group element that maps $\Sigma$ into the group $G$.
We shall work with complex coordinates
$z,\bar{z}$ and choose the  metric $h_{z\bar{z}} =h^{z\bar{z}}=1$.
The supersymmetric extension $I(g,\Psi)$ of the above action is obtained by
adding to it
the free action of Weyl fermions $\Psi_{L}$ and $\Psi_{R}$ in the
complexification of the
adjoint representation of $G$
\cite{EW92}:\footnote{We shall restrict ourselves to models that have equal
number of supersymmetries in both the left and the right moving sectors.}
\eq
I(g,\Psi) =I(g) + \frac{i}{4\pi} \int d^2z Tr(\Psi_L \partial_{\bar{z}} \Psi_L
+
\Psi_{R} \partial_z \Psi_R )
\en
Under the supersymmetry action the fields transform as follows
\eqn
\delta g& =& i \epsilon_- g \Psi_L +i \epsilon_+ \Psi_R g  \nn
&& \nn
\delta \Psi_L& =& \epsilon_- (g^{-1} \partial_z g -i \Psi_{L}^{2})  \nn
&& \nn
\delta \Psi_R& =& \epsilon_+ (\partial_{\bar{z}} g g^{-1} + i \Psi_{R}^{2} )
\enn
To gauge a diagonal subgroup $H$ of the $G_L\times G_R$ symmetry of the WZW
model
one introduces gauge fields $(A_z,A_{\bar{z}})$ belonging to the subgroup $H$.
 The gauge invariant action, which does not involve any kinetic
energy term for the gauge fields,  can be written as:
\eq
I(g,A) = I(g) + \frac{1}{2\pi} \int_{\Sigma} d^2z Tr(A_{\bar{z}} g^{-1}
\partial_z g - A_z \partial_{\bar{z}}g g^{-1} +A_{\bar{z}}g^{-1} A_{z}g
-A_{\bar{z}}
A_z)
\en
 The gauge transformations of the fields are defined as \cite{EW92}:
\eqn
\delta g& =& [u,g] \nn
\delta A_i& =& -D_iu = -\partial u -[A_i,u]
\enn
Let ${\cal G}$ and ${\cal H}$ be the complexifications the Lie algebras of $G$
and $H$. Then ${\cal G}$ has an orthogonal decomposition
\eq
{\cal G} = {\cal H} \oplus {\cal T}
\en
where ${\cal T}$ is the orthocomplement of ${\cal H}$.

To supersymmetrize the gauged WZW model one introduces Weyl fermions with
values in ${\cal T}$ minimally coupled to the gauge fields and otherwise free
\eq
I(g,A,\Psi) = I(g,A) + \frac{i}{4\pi} \int d^2z Tr(\Psi_L D_{\bar{z}} \Psi_L
+ \Psi_R D_z \Psi_R )
\en
It is invariant under the supersymmetry transformation laws:
\eqn
\delta g& =& i \epsilon_- g \Psi_L +i \epsilon_+ \Psi_R g \nn
&& \nn
\delta \Psi_L& =& \epsilon_- (1-\Pi) (g^{-1} D_z g -i \Psi_L^2 ) \nn
&& \nn
\delta \Psi_R& =& \epsilon_+ (1- \Pi) (D_{\bar{z}} g g^{-1} +i \Psi_R^2)  \nn
&& \nn
\delta A& =& 0
\enn
where $\Pi$ is the orthogonal projection of ${\cal G}$ onto ${\cal H}$.

Witten showed that when the coset space $G/H$ is Kahler the above action has
$N=2$ supersymmetry. In the Kahler case ${\cal T}$ has a decomposition
\eq
{\cal T} = {\cal T}_+ \oplus {\cal T}_-
\en
where ${\cal T}_+$ and ${\cal T}_-$ are in complex conjugate representations of
$H$.
Then the action can be written in the form \cite{EW92}
\eq
I(g,\Psi,A) = I(g,A) + \frac{i}{2\pi} \int d^2z Tr ( \beta_L D_{\bar{z}}
\alpha_L + \beta_R D_z \alpha_R )
\en
where
\eqn
\alpha_L& =& \Pi_+ \Psi_L  \nn
 \beta_L& =& \Pi_- \Psi_L \nn
\alpha_R& =& \Pi_+ \Psi_R \nn
\beta_R& =& \Pi_- \Psi_R
\enn
with $\Pi_+$ and $\Pi_-$ representing the projectors onto the subspaces
${\cal T}_+$ and ${\cal T}_-$.
Denoting the chiral and anti-chiral supersymmetry generators in left  and right
moving
sectors as $G_L , \bar{G}_L$ and $G_R , \bar{G}_R$ , respectively,  let us
define
the operators
\eqn
\delta_G& =& \epsilon_- G_L + \epsilon_+ G_R \nn
&& \nn
\bar{\delta}_G& =& \epsilon_- \bar{G}_L + \epsilon_+ \bar{G}_R
\enn
where the $\epsilon_{\pm}$ are  anticommuting Grassmann parameters.

Their actions on the fields of the theory are
\eqn
\delta_G g& =& i \epsilon_- g \alpha_L + i \epsilon_+ \alpha_R g \nn
&& \nn
\delta_G \alpha_+& =&-i \epsilon_- \Pi_+ \alpha_L^2  \nn
&& \nn
\delta_G \alpha_-& =& i \epsilon_+ \Pi_- \alpha_R^2 \nn
&& \nn
\DG \B_L& =& \E_- \Pi_- (g^{-1} D_z g -i \B_L \A_L -i \A_L \B_L) \nn
&& \nn
\DG \B_R& =& \E_+ \Pi_+ (D_{\bar{z}} g g^{-1} + i \B_R \A_R +i \A_R \B_R)
 \nn
&& \nn
\DGB g& =& i\E_- g \B_L +i \E_+ \B_R g  \label{eq:ME} \\
&& \nn
\DGB \B_L& =& -i \E_- \Pi _- \B_L^2  \nn
&& \nn
\DGB \B_R& =& i\E_+ \Pi_+ \B_R^2  \nn
&& \nn
\DGB \A_L& =& \E_- \Pi_+ ( g^{-1} D_z g -i \B_L \A_L -i \A_L \B_L  ) \nn
&& \nn
\DGB \A_R& =& \E_+ \Pi_- (D_{\bar{z}} g g^{-1} + i \A_R \B_R +i \B_R \A_R)
\nonumber
\enn
To prove that the action is invariant under these transformations one needs
to use the facts that
\eqn
 {\rm Tr} ab& =& 0   \leftrightarrow   a,b \in {\cal T}_+ \nn
 {\rm Tr} ab& =&0  \leftrightarrow   a,b \in {\cal T}_-   \nn
{[} {\cal T}_+, {\cal T}_+ {]}& \subset& {\cal T}_+  \nn
{[} {\cal T}_-, {\cal T}_-{]}& \subset & {\cal T}_-
\enn

Since the theory is known to be conformally invariant it suffices to restrict
oneself
to checking the global supersymmetries in order to prove $N=2$ superconformal
invariance. Using the equations of motion it is straightforward to show that
\eqn
{[}\DG , \DGB^{\prime}{]}& =& i \E^{\prime}_- \E_- D_z + i  \E_+^{\prime} \E_+
D_{\bar{z}} \nn
&& \nn
{[} \DG , \DG^{\prime} {]}& =& 0  \nn
&& \nn
{[}\DGB , \DGB^{\prime} {]}& =& 0
\enn
or equivalently
\eqn
\{G_L,\bar{G}_L\}&=&-iD_z \nn
&& \nn
\{G_R,\bar{G}_R\}&=&-iD_{\bar{z}}  \label{eq:AR} \\
&& \nn
\{G_L,G_L\}&=&\{\bar{G}_L,\bar{G}_L\}=0 \nn
&& \nn
\{G_R,G_R\}&=&\{\bar{G}_R,\bar{G}_R\}=0 \nonumber
\enn

That the supersymmetry algebra closes only on-shell is expected since one is
working
in component formalism.

\section{N=4 Supersymmetric Gauged WZW  Models}
\setcounter{equation}{0}
We saw above that the existence of a second supersymmetry in a supersymmetric
gauged WZW model is guaranteed when the coset space $G/H$ is Kahlerian , i.e it
admits a complex structure. Therefore one would expect that to have $N=4$
supersymmetry one needs coset spaces with three complex structure which
anti-commute with each other
and form a closed algebra. This is also expected from the study of $N=4$
supersymmetric  sigma models \cite{AF,SSTP1} and the unitary realizations of
$N=4$  superconformal algebra over quaternionic symmetric spaces
\cite{SSTP2,AV,GH91,GH92}.
In section we have reformulated the construction of \cite{GH91,GH92} for $N=2$
superconformal algebras that are extendable to $N=4$ SCA's so as to be able
to relate them to gauged supersymmetric WZW models. Let us now show that the
$N=1$
supersymmetric WZW models based on the groups $G\times U(1)$ of section 3 with
a gauged subgroup $H$ such that $G/H\times SU(2)$ is a quaternionic symmetric
space
have actually $N=4$ supersymmetry.

We shall designate the generators of $G\times U(1)$ as we did in sections 2 and
3
,i.e $K_0, K_1,K_2,K_3,U_a,U^a$ and $H^b_a$ , where $K_0$ is the generator of
the  additional
$U(1)$ factor. We normalize the generator $K_0$ such that
\eq
Tr K_0^2 =Tr K_1^2=Tr K_2^2 =Tr K_3^2
\en
 The fermions associated with the grade $\pm1$ subspaces of the Lie
algebra of $G$ will be denoted as $\psi_a,\psi^a$ as before. However, the
fermions
associated with $K_0$ and $K_i$
will be denoted as $\xi^0, \xi^i (i=1,2,3)$. Then the fermions
in the coset $G\times U(1)/H$ can be written as
\eq
\Psi = 2K_0 \xi^0 + 2K_i \xi^i + U_a \psi^a + U^a \psi_a
\en
for both the left and the right moving sectors.
The coset $G\times U(1)/H$ can be given a Kahler decomposition so that we can
write $\Psi$ as

\eq
\Psi= \A + \B
\en
where
\eqn
\A& =& U_a \psi^a + K_+ (\xi^1-i\xi^2) + (K_3+iK_0)(\xi^3-i\xi^0) \nn
&& \nn
\B& =& U^a \psi_a + K_- (\xi^1 +i \xi^2) + (K_3-iK_0)(\xi^3+i\xi^0)
\enn
The complex structure $C_3$ corresponding to this decomposition acts on $\Psi$
as
follows
\eq
C_3 \Psi = -i \A +i\B
\en
where the index 3 in $C_3$ signifies the fact in the subspace
$G/H \times SU(2)$ its action corresponds to commutation with$-iK_3$.
Similarly, we can give a Kahler decomposition of  the coset space $G\times
U(1)/H$
which selects out $K_1$ or $K_2$. For $K_1$ we have:
\eqn
\Psi& =& \A^{\prime} + \B^{\prime} \nn
&& \nn
\A^{\prime}& =& U_a^{\prime} \psi^{a^{\prime}}
 +(K_2+iK_3)(\xi^2-i\xi^3) + (K_1+iK_0)(\xi^1-i\xi^0)  \nn
&& \nn
\B^{\prime}& =& U^{a^{\prime}} \psi_a^{\prime} +(K_2-iK_3)(\xi^2+i\xi^3)
+ (K_1-iK_0)(\xi^1+i\xi^0)
\enn
Under the action of the corresponding complex structure $C_1$ we have
\eq
C_1 \Psi = -i\A^{\prime} +i\B^{\prime}
\en
In the case of $K_2$ we have
\eqn
\Psi&=& \A^{\prime \prime} +\BPP \nn
&& \nn
\APP& =&  U_a^{\prime \prime}\psi^{a^{\prime \prime}}
 +(K_3+iK_1)(\xi^3-i\xi^1) + (K_2+iK_0)(\xi^2-i\xi^0) \nn
&& \nn
\BPP& =& U^{a^{\prime \prime}} \psi_a^{\prime \prime}
+ (K_3-iK_1)(\xi^3+i\xi^1) +(K_2-iK_0)(\xi^2+i\xi^0)
\enn
with the complex structure action
\eq
C_2 \Psi = -i \APP +i\BPP
\en
The fermionic part of the action can then be written in three different
ways involving the pairs $(\A,\B),(\AP,\BP)$ and $(\APP,\BPP)$.
\eqn
I(\Psi,A)&=& \frac{i}{4\pi}\int d^2z Tr(\Psi_L D_{\bar{z}}\Psi_L
+ \Psi_R D_z \Psi_R ) \nn
&& \nn
&=& \frac{i}{2\pi}\int d^2z Tr( \B_L D_{\bar{z}}\A_L +
\B_R D_z \A_R) \nn
&& \nn
&=& \frac{i}{2\pi}\int d^2z Tr(\BP_L D_{\bar{z}} \AP_L +
\BP_R D_z \AP_R) \nn
&& \nn
&=& \frac{i}{2\pi} \int d^2z Tr(\BPP_L D_{\bar{z}} \APP_L +
\BPP_R D_z \APP_R)
\enn
Because of the presence of gauge covariant derivatives in the action the above
relations may not appear obvious. However, the fact that the fermions $\xi^0$
and $\xi^i$ are singlets of the gauge group and the fact that the complex
structures' actions on the coset space $G/H \times SU(2)$ generators commute
with the gauge group $H$ action imply their validity.     For each form of the
action in terms of an $(\A,\B)$ pair one can define a pair of supersymmetry
transformations in each sector as in equations  \ref{eq:ME} of the previous
section. Let us denote them as $(G,\bar{G})
,(G^{\prime}.\bar{G}^{\prime})$ and $(G^{\prime \prime},\bar{G}^{\prime
\prime})$:
\eqn
(G,\bar{G}) &\leftrightarrow & (\A,\B) \nn
&& \nn
(G^{\prime}, \bar{G}^{\prime}) & \leftrightarrow &
(\AP, \BP) \nn
&& \nn
(G^{\prime \prime}, \bar{G}^{\prime \prime}) & \leftrightarrow &
(\APP,\BPP)
\enn
Each pair of these operators in both sectors satisfy the $N=2$ supersymmetry
algebra given by the equations \ref{eq:AR}.   However they are not all
independent. The sum of each pair gives the manifest $N=1$ supersymmetry
generator
of the model in both sectors, which we shall denote as $G^0$:
\eq
G^0 =\frac{1}{\sqrt{2}}(G+\bar{G}) = \frac{1}{\sqrt{2}}(G^{\prime}+
\bar{G}^{\prime})= \frac{1}{\sqrt{2}}(G^{\prime \prime}+ \bar{G}^{\prime
 \prime})
\en
They satisfy
\eqn
\{G^0_L,G^0_L\}&=& -iD_z \nn
&& \nn
\{G^0_R,G^0_R\}&=&-iD_{\bar{z}}
\enn
We define three additional supersymmetry generators (in each sector)
\eqn
G^3 &=& \frac{1}{i\sqrt{2}}(G-\bar{G}) \nn
&& \nn
G^1 &=& \frac{1}{i\sqrt{2}}(G^{\prime}-\bar{G}^{\prime}) \nn
&& \nn
G^2 &=& \frac{1}{i\sqrt{2}}(G^{\prime \prime}
-\bar{G}^{\prime \prime})
\enn
 Each one of these three supersymmetry generators anticommute with $G^0$ and
obey
\eqn
\{G^3_L,G^3_L\}&=&\{G^1_L,G^1_L\}=\{G^2_L,G^2_L\}=-iD_z  \nn
&& \nn
\{G^3_R,G^3_R\}&=&\{G^1_R,G^1_R\}=\{G^2_R,G^2_R\}=-iD_{\bar{z}} \nn
&& \nn
\{G^0_L,G^1_L\}&=&\{G^0_L,G^2_L\}=\{G^0_L,G^3_L\}=0 \nn
&& \nn
\{G^0_R,G^1_R\}&=&\{G^0_R,G^2_R\}=\{G^0_R,G^3_R\}=0
\enn
as a consequence of the equations \ref{eq:AR}.
To prove that $G^{\mu} (\mu=0,1,2,3)$
form an $N=4$ superalgebra we need to
further show that the $G^i (i=1,2,3)$ anticommute with each other. Direct
proof using their actions on the fields of the theory  is quite involved.
However there is a simple proof which uses the quaternionic structure
of the coset $G\times U(1)/ H$. We first note that the complex structures
$C_i$ obey the relation
\eq
C_i C_j = C_k
\en
where $i,j,k$ are in cyclic permutations of $(1,2,3)$. Furthermore, the
action is invariant under the replacement of $\Psi$ by $C_i \Psi$.
If we start from an action with $\Psi$ replaced by ,say, $C_1 \Psi $
then the
manifest $N=1$ supersymmetry will be generated by $G^1$ and the second
supersymmetry generated by the Kahler decomposition with respect to the
complex structure $C_3$ will be $G^2$ since $C_3 C_1=C_2$. Hence by the
results of the previous section we have
\eq
\{G^1,G^2\}=0
\en
and by cyclic permutation we have
\eq
\{G^2,G^3\}=\{G^3,G^1\}=0
\en
To summarize we have shown that the four supersymmetry generators
$G^{\mu}$ satisfy the $N=4$ supersymmetry algebra:
\eqn
\{ G^{\mu}_L, G^{\nu}_L\}& =& -i \delta^{\mu \nu} D_z \nn
&& \nn
\{ G^{\mu}_R , G^{\nu}_R\}& = & -i \delta^{\mu \nu} D_{\bar{z}}\nn
&& \nn
\mu,\nu,...&=&0,1,2,3
\enn
Since the gauged $WZW$ models considered above
are known to be conformally invariant we have
thus proven that they are invariant under $N=4$ superconformal
transformations.

If we restrict ourselves to the case $G=SU(2)$ then we have the supersymmetric
WZW model on the group manifold $SU(2)\times U(1)$ with no gauge fields. The
$N=4$
supersymmetry of this model was studied in \cite{ARSS} using superspace
techniques.
In the general case, we can integrate out the four fermions $\xi^a$ that do not
couple
to the gauge fields of $H$ as well as the Abelian $U(1)$ field generated by
$K_0$ .
Then we expect the resulting theory  to be invariant under the non-linear $N=4$
SCA
${\tilde{\cal A}}_{\gamma}$.

 The quantization of gauged WZW models is known to yield  the coset space
realizations of conformal algebras \cite{GWZW,EW92b}.
 Therefore the quantization of the
$N=4$ gauged WZW models we have given above  is expected to give
the unitary realizations
of maximal $N=4$  SCA's with $SU(2)\times SU(2) \times U(1)$ symmetry
that were presented in section 3.  Their quantization
and the study of more general gauged WZW models
with  extended non-linear superconformal symmetry will be left to
future investigations. \\
\  \\

{\it Acknowledgements}: Useful discussions with Edward Witten and
Michael Douglas are gratefully acknowledged.

\end{document}